\documentclass[showpacs, preprintnumbers, twocolumn]{revtex4}
\usepackage{dcolumn}
\usepackage{bm}
\usepackage{graphicx}

\begin{document} 

\title{Anchored Critical Percolation Clusters and 2-D Electrostatics}

\begin{abstract}
We consider the densities of clusters, at the percolation point of a two-dimensional system, which are anchored in various ways to an edge.   These quantities are calculated by use of conformal field theory and computer simulations.  We find that they are given by simple functions  of the potentials of  2-D electrostatic dipoles, and that a kind of superposition {\it cum} factorization applies.  Our results broaden this connection, already known from previous studies, and we present evidence that it is more generally valid. An exact result similar to the Kirkwood superposition approximation emerges.


\end{abstract}

\author{P. Kleban}
\email{kleban@maine.edu}
\author{J. J. H. Simmons}
\email{Jacob.Simmons@umit.maine.edu}
\affiliation{LASST and Department of Physics and Astronomy, University of Maine, Orono ME 04469}
\author{R. M. Ziff}
\email{rziff@engin.umich.edu}
\affiliation{Michigan Center for Theoretical Physics and Department of Chemical Engineering, University of Michigan, Ann Arbor MI 48109-2136}
\maketitle

Percolation in two-dimensional systems is an area with a long history, which remains under very active current study. A plethora of methods has been applied to critical 2-D percolation, including  conformal field theory (CFT)  \cite{JCxing}, modular forms \cite{KZ}, computer simulation \cite{KZi}, other field-theoretic methods \cite{BD}, Stochastic L\"owner Evolution (SLE) processes \cite{JD} and other rigorous methods \cite{MA}.  (The literature is so extensive that we have cited only a very few representative works.)  

It has long been known that the behavior of percolating systems at or near the percolation point is intimately connected with two-dimensional electrostatics, or more precisely the Coulomb gas (Gaussian model) of field theory.  In particular, Coulomb gas methods have given many important results, for instance crossing formulas on a torus \cite{P}.  Recently, Smirnov \cite{S} has proved that the crossing probability in critical percolation in an equilateral triangle may be described as the boundary value of a certain very simple problem in 2-D electrostatics.  In this Letter, we deepen and extend this connection by demonstrating that certain correlation functions, which specify the density of clusters at the percolation point, are given as simple functions of the potentials of two-dimensional dipoles.  

In addition we show that a kind of superposition applies.  An exact result resembling the Kirkwood superposition approximation emerges. 

We find indications that our results also apply when the anchor points are not on the edge, and hold for the critical Potts models. 

Our results are found via CFT  \cite{BPZ} with verification by   computer simulations.  

Our work considers the (average) density of anchored critical clusters in 2-D percolation. The clusters are in the upper half-plane, and are constrained to touch the real axis at a specified point or points, or in an interval between two given points. By conformal invariance, the corresponding results for any compact region may be obtained by using the appropriate transformation.  The related problem of the density of clusters in statistical systems in other geometries with fixed boundaries has been investigated in \cite{RS}.  

The first case we consider anchors the clusters at a single point, which we may take as the origin. To find their density, we use  Cardy's CFT analysis  \cite{JCxing} of  crossing probabilities in critical 2-D percolation for systems with an edge. This approach makes use of the Q-state Potts model and the boundary operator $\phi_{1,2}(x)$, which changes the boundary conditions from fixed to free at a point $x$ on the real axis. For percolation ($Q \to 1$) it has conformal dimension $h_{1,2}=0$.  Two such boundary operators  $\phi_{1,2}(x_1)\phi_{1,2}(x_2)$ may be used to count the number of clusters touching the real axis for $x_1 < x < x_2$ \cite{JCJap}.  

If we bring $x_1$ and $x_2$ together, the operator product expansion (OPE) is applicable.  Two terms arise.  One involves the unit operator, which has conformal dimension $h_{1,1}=0$, and the other $\phi(x)=\phi_{1,3}(x)$.  This latter operator has conformal dimension $h_{\phi}=h_{1,3}=1/3 > 0$, and thus (since the corresponding term vanishes in this limit) represents the clusters anchored between the points.  (Since the boundary spins are fixed, in this limit there must be exactly one anchored cluster, as was also argued in \cite{JCxing2}.)  The dimension $1/3$ also governs the vanishing of the crossing probability for, e.g., a long thin rectangle.  The other term, from the unit operator, represents clusters that are anchored on the real axis everywhere {\it except} at $x_1 < x < x_2$.  In this case the density becomes independent of $x$, as it should.

The operator $\psi(z)$, which measures the cluster density at a point $z$, has dimension $h_{\psi} = 5/96$.  The corresponding conformal operator is identified as $\psi=\phi_{3/2,3/2}$.  This is the $Q\to 1$ ($m \to 2$) limit of the magnetization operator $\phi_{(m \pm 1)/2,(m+1)/2}$ for the $Q$-state Potts model \cite{DF,JCedge}.

The cluster density with one anchor point is therefore given by
\begin{eqnarray} \label{ptdens}
\rho_{pt}(z) &=& \langle\phi(0) \psi(z)\rangle_{1/2}\nonumber \\
        &=& \langle\phi(0) \psi(z) \psi(\bar z)\rangle  \nonumber \\
	  &=& \frac 1{z^{h_{\phi}} \bar z^{h_{\phi}} (z-\bar z)^{2 h_{\psi}-h_{\phi}}},
\end{eqnarray}
where $\langle\cdots\rangle_{1/2}$ and $\langle \cdots\rangle$ denote half-plane and full plane expectation values respectively, the second line follows because half-plane correlation functions are given by full-plane correlators with ``image" operators \cite{JCedge}, and the third line makes use of the standard form for the three-point function.  (Here and below we ignore certain multiplicative constants.)  

Simplifying and introducing polar coordinates gives 
\begin{eqnarray} \label{ptdens2}
\rho_{pt}(z) &=& \frac{(2y)^{11/48}}{r^{2/3}}\,\nonumber \\
	  &=& \frac {(2\sin \theta)^{11/48}} {r^{7/16}}.  
\end{eqnarray}

It follows that $\rho_{pt} \to 0$ when $z$ approaches the real axis or $\infty$. It vanishes as $\rho_{pt}  \sim y^{-7/16}$ when $y = \Im(z) \to \infty$, and as $y^{11/48}$ when $y \to 0$  (for $x\ne0$), so that the density has a maximum as a function of $y$.  An exception occurs for the limit $z \to 0$, where $\rho_{pt}$  diverges as $r^{-7/16}$.  

An infinite value for the density in a real system is of course not possible. Its value on any lattice, for instance, will be bounded via a cut-off involving the lattice constant.

On the other hand,  (\ref{ptdens}) may be expressed as
\begin{equation} \label{ptdens3}
\rho_{pt}(z) =\frac 1{y^{5/48}}(\Phi_{dip}(z;0))^{1/3},
\end{equation}
where $\Phi_{dip}(z;x_0)=2y/((x-x_0)^2+y^2)$ may be taken as either the real (physical) part of the complex potential of an 2-D unit point dipole in the $y$ direction located at position $x_0$ on the real axis  or the imaginary part when the dipole is in the $-x$ direction.

Note the contours  $\Phi_{dip}(z;x_0) = a$ are circles of radius $1/a$, centered at $(x_0, y_0 = 1/a)$ and all tangent to the real axis at $x_0$.  Hence the contours of constant $y^{5/48} \rho_{pt}$ are likewise circles.  The contours of $\rho_{pt}$ itself have the form of circles deformed by pushing against the real axis.  Thus lines parallel to the $y$-axis pass ``over the shoulder" of the density and have a maximum (when $x \ne 0$).

The above results are a nice example of the power of conformal invariance.  Up to a multiplicative constant, the density $\rho_{pt}(z)$ is completely determined by only two numbers, the conformal dimensions $1/3$ and $5/96$. 

Functions of the same form as (\ref{ptdens}) arise generally in the related problem of the evaluation of two-point correlation functions of a single operator in the upper half-plane when one of the operators approaches the real axis \cite{JCedge}.

Comparison of  (\ref{ptdens3}) with computer simulations is excellent, as described in detail below.

It is interesting to note that (\ref{ptdens3}), when transformed via a conformal map $w=w(z)$, preserves its form.  It remains a product of two factors.  The first is a function of $w$ only, the form of which depends on the size and shape of the new region, but is {\it independent} of the anchoring point $w(0)$.  This is multiplied by $\tilde \Phi_{dip}(w;w(0))^{1/3}$, the (real or imaginary) potential of a dipole at the mapped anchor point  $w(0)$ in the new region, with grounded boundary.

The next case of interest has the cluster anchored at the two points $x_a$ and $x_b = x_a + D$.  By the arguments above, the density with two anchor points is
\begin{equation} \label{2ptdens}
\rho_{2pts}(z;x_a,x_b) =  \langle\phi(x_a) \phi(x_b) \psi(z) \psi(\bar z)\rangle.\label{2ptsdens}
\end{equation}
Following the standard treatment (see \cite{BPZ} or  \cite{JCedge}), we rewrite the correlation function in (\ref{2ptsdens}) so that 
\begin{equation} \label{2ptdens2}
\rho_{2pts}(z;x_a,x_b) =(x_b-x_a)^{-2 h_\phi} (z-\bar z)^{-2 h_\psi} F(\eta),
\end{equation}
where the cross-ratio $\eta=[(x_b-x_a)(\bar z-z)]/[(z -x_a)(\bar z-x_b)]$.  

Since $\phi = \phi_{1,3}$ is a level-three operator, by standard CFT analysis, $F(\eta)$ satisfies the third-order differential equation 
\begin{eqnarray} 
 \left[108 \eta^2 (\eta-1)^3 \frac{d^3}{d \eta^3}+72 (\eta-1)^2 \eta (5 \eta-1) \frac{d^2}{d \eta^2} \right. \nonumber \\
\label{2ptde}
\left. 3(\eta-1)(35\eta^2-24)\frac{d}{d \eta}+5\eta(\eta-2)\right]F(\eta)=0. 
\end{eqnarray} 

Since percolation is a logarithmic CFT, the use of a differential equation like (\ref{2ptde}), which arises from a Virasoro null vector, can be problematical.  However this appears not to be the case here (see \cite{EF} and  \cite{LG}).

We can find the solution to (\ref{2ptde}) of interest by making use of the OPE of $\phi(x_a) \phi(x_b)$ as $x_a \to x_b$.  This has three  terms, involving either the unit operator, $\phi(x_a)$ or $\phi_{1,5}(x_a)$.  On the other hand, $\rho_{2pts}(z;x_a,x_b)$ should be proportional to $\rho_{pt}(z;x_a)$ in this limit.  The first OPE term clearly gives a contribution which is independent of $x_a$ and thus may be discarded.  The operator $\phi_{1,5}$ has been argued to create two clusters separated by a dual line \cite{JCxing2}, and thus cannot contribute to $\rho_{pt}$.
What remains may be expressed in the interesting factorized form 
\begin{eqnarray}  \label{2ptdens3}
\rho_{2pts}(z;x_a,x_b)&=&  \\ 
 &&\frac 1{y^{5/48}}\frac 1{D^{1/3}} (\Phi_{dip}(z;x_a) \Phi_{dip}(z;x_b))^{1/6}. \nonumber
\end{eqnarray}

Clearly (\ref{2ptdens3}) reduces to (\ref{ptdens3}), as expected,  if we take $D \to 0$ (and set $x_a = 0$). Note also that in the limit $z \to x_a$ or $z \to x_b$,  $\rho_{2pts}=1/D^{2/3}$, which is just the (unnormalized) probability of a cluster connecting  $x_a$ and $x_b$. (The latter limit also implies that $\psi$ reduces to $\phi$ on a free boundary, as discussed further below.)

Multiplying by $y^{5/48}$ and comparing (\ref{ptdens3}) and (\ref{2ptdens3}) we see that the logarithm of the latter is linear in the logarithm of the former.  Thus a kind of superposition applies. 

\begin{figure}[h]
\begin{center}
\includegraphics[width=3.0in]{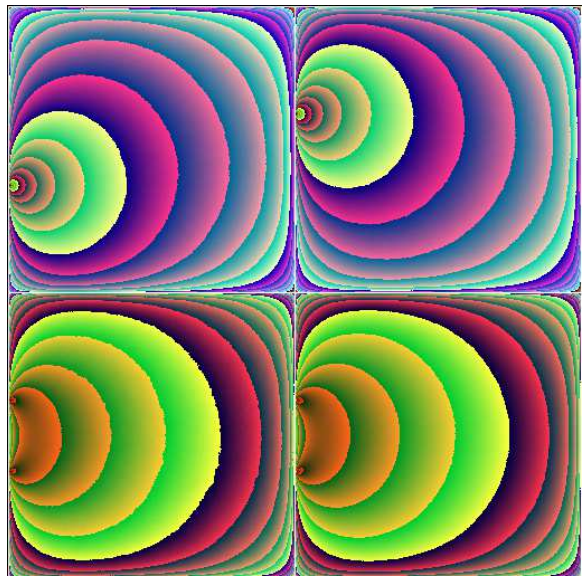}
 \end{center}
\caption{(color online) Simulation results for $\rho_{pt}(z;y)$ and $\rho_{2pts}(z;y_a,y_b)$.} \label{sim}
\end{figure}

To test these predictions, we carried out simulations on square  lattices of sizes
$255\times 255$, $511\times 511$, $1023\times 1023$, and $2047\times 2047$ sites using bond percolation at the critical point $p = 1/2$,
and keeping track of the wetted sites connected to the two
anchor points, positioned $3/8$ and $5/8$ the way up the $y$-axis
(e.\ g., in the $255\times 255$ case, the two points separated by $\pm 32$ lattice spacings from the center).  All boundaries were free.  We used a cluster
growth algorithm starting from either anchor point, and if the
cluster touched the other anchor point, we also averaged that
density separately.  Thus, we found the density of clusters touching
either one or both anchor points.  In Fig. \ref{sim}, the upper left panel
shows the  density of points touching the lower anchor point,
and the upper right shows the density of clusters touching the
upper anchor point. These densities are normalized to be one at the anchor point.  The lower left panel shows the density of
clusters touching both anchor points, normalized similarly so that it is one at either anchor point.  In the lower
right panel, we show the square root of the product of the two simulated
one-point densities divided by the probability that the two anchors
are connected together, finally multiplied by a constant $C$ to make
the contours agree quantitatively with those
of the lower left panel.  The effective value of $C$ is one when $z$ is at either anchor point, by definition. However, it changes to $C=1.030 \pm 0.001$ within a few lattice spacings, independent of lattice size.  We interpret $C$ below. 

We find quantitatively similar results for site percolation on square and triangular lattices, and bond percolation on a triangular lattice.

We made an additional run of $10^8$ samples and compared the
density along the horizontal centerline ($y = 1/2$), and found
agreement with (\ref{2ptdens3}) within $1\%$, with slightly larger deviations
at the two endpoints (boundaries).  We find similar agreement of the simulated one point density with (\ref{ptdens3}).

Thus the factorization in (\ref{2ptdens3}) agrees with simulation results, supporting our arguments for discarding the solution of (\ref{2ptde}) giving rise to the $\phi_{1,5}$ term.  We also find numerically that this factorization holds asymptotically (with the same value of $C$) when one or both anchor points is not on the edge. (In this case the change of $C$ from $1$ to the quoted value occurs over a number of lattice spacings that scales with lattice size, indicating that the rhs of (\ref{2ptdens3}) has an additional term.) Furthermore, CFT calculations indicate that a similar factorization applies to the critical Potts models as well.  Hence these results seem to apply more generally. 

Note that CFT implies that the factorization in (\ref{2ptdens3}) remains valid under any conformal mapping.

To understand these results better we combine (\ref{ptdens3}) and (\ref{2ptdens3}), which, in an obvious notation, gives
\begin{equation} \label{gm}
\rho_{2pts}(z;x_a,x_b) = \frac 1{D^{1/3}} \sqrt{\rho_{pt}(z;x_a) \rho_{pt}(z;x_b)}. 
\end{equation}
Now the densities in (\ref{gm}), (and the factor $1/D^{1/3}$), may be understood as the probability ${\cal P}$ of a cluster that connects the points in question (or the anchor points).  Now since, as mentioned, $\psi$ reduces to $\phi$ on a free boundary, if we take $z \to x_c$, (\ref{gm}) reduces to 
\begin{eqnarray} \label{const}
&&\langle\phi(x_a) \phi(x_b) \phi(x_c)\rangle= \\
&&C\sqrt{\langle\phi(x_a) \phi(x_b) \rangle \langle\phi(x_a) \phi(x_c)\rangle\langle \phi(x_b) \phi(x_c)\rangle}, \nonumber
\end{eqnarray}
where $C$ is exactly the (boundary) OPE coefficient of the term that we retained in the solution of (\ref{2ptde}).  Therefore
\begin{equation} \label{probs}
{\cal P}(z,x_a,x_b) = C\;\sqrt{{\cal P}(x_a,x_b){\cal P}(z,x_a){\cal P}(z,x_b)},
\end{equation}
with $C$ as above.  Except for the square root, (\ref{probs}) resembles the Kirkwood superposition approximation familiar from the theory of fluids \cite{K}, which has been applied to percolation \cite{GK}. However it should be emphasized that (\ref{probs}) is both exact and universal.

Finally we consider a cluster anchored along the entire interval $x_1 \le x \le x_2$. Arguing as  above,  the cluster density is now given by the four-point correlation function 
\begin{equation} \label{intdens}
\rho_{int}(z;x_1,x_2) = \langle\phi_{1,2}(x_1) \phi_{1,2}(x_2) \psi(z) \psi(\bar z)\rangle. 
\end{equation}
Proceeding as for (\ref{2ptdens}), and redefining $\eta$ by replacing $x_a \to x_1$ and $x_b \to x_2$, gives 
\begin{eqnarray} \label{intde}
&&\left[\; \eta (1-\eta)^2 \frac{d^2}{d \eta^2} \right. \\
&&\left.+\frac23 (1-\eta)(1-2 \eta) \frac{d}{d \eta}- \frac23 \, h_{\psi} \, \eta \;\right] F(\eta)=0. \nonumber
\end{eqnarray} 
Since $\phi_{1,2}$ is a level-two operator (\ref{intde}) is second order.

The limit $x_2, x_1 \to 0$ gives a single cluster anchored at the origin. The solution of (\ref{intde}) goes either as $F_- \sim \eta^{1/3}$ or as $F_+ \sim \eta^{0}$ in this limit, corresponding, respectively, to the appearance of $\phi$ or of the unit operator in the OPE, as mentioned above.  Retaining $F_-$ only, we find that (\ref{intsol}) (see below)  reproduces (\ref{ptdens}), as it should.  

Note that setting $h_{\psi}=0$ reduces (\ref{intde}) to Cardy's differential equation for the crossing probability \cite{JCxing}, as it should, since that quantity involves four boundary operators $\phi_{1,2}$ with dimension zero.

The two independent solutions $F_+$ and $F_-$ of (\ref{intde})  may be written as
\begin{equation} \label{desols}
F_{\pm}(\eta) = \left(\frac{2-\eta}{2 \sqrt{1 - \eta}} \pm 1\right)^{1/6}.
\end{equation}

We next set $z=x+iy$ and $x_1=-L/2, x_2=L/2$ so that the cluster is anchored on an interval of length $L$ centered at the origin. Substituting these values into the cross-ratio $\eta$, one finds, with appropriate choice of the square root,
\begin{eqnarray} \label{intsol}
&&\rho_{int}(z;L)=\frac1{y^{5/48}}  \Bigg(1+ \\
&& \frac{L^2-4(x^2+y^2)}{\sqrt{L^4-8L^2(x^2-y^2)+16(x^2+y^2)^2}}\Bigg)^{1/6}. \nonumber
\end{eqnarray}

Note that for $L \to 0$, $\rho_{int}(z;L) \to (L^{1/3}/2^{1/6})\;\rho_{pt}(z)$, as expected since the interval reduces to a point.

In the limit $y \to 0$,  (\ref{intsol}) becomes a power series in $y^2$ when $|x| < L/2$ and $y^{1/3}$ times a  power series in $y^2$ when $|x| > L/2$. This shows that in the boundary OPE \cite{DD,CL}  for $\psi$, to leading order an operator of dimension $h=0$ (the unit operator, presumably) appears for fixed boundary conditions while an operator of dimension $h=1/3$ ($\phi$, by the result above) appears for free boundary conditions.  The latter conclusion emphasizes again that in order for $\psi$ to measure the density  it must connect to a cluster.  Note that an analogous dependence of the boundary OPE on boundary conditions occurs for the magnetization operator ($\phi_{2,2}$ or $\phi_{1,2}$) in the Ising model \cite{CL}.

In order to make contact with 2-D electrostatics again, we first integrate $\Phi_{dip}$ over the position of the dipole.  Explicitly, let 
$\Psi(z;L) = \int_{-L/2}^{L/2}\Phi_{dip}(z;x_0)d x_0$. Note that if $\Phi_{dip}$ is taken as the imaginary potential as above, then $\Psi(z;L)$ is the imaginary potential for a  {\it finite} dipole from $-L/2$ to $L/2$ (in the other interpretation it is the potential of a dipole layer between these points). It is straightforward to show that (\ref{intsol}) may be expressed as 
\begin{equation} \label{intdens}
\rho_{int}(z;L)=\frac 1 {y^{5/48}}\; \sin^{1/3}\left(\frac{\Psi(z;L)}{4}\right).
\end{equation}
Again we see in (\ref{intdens}) a kind of exponential additiveness, via the sine function.

In summary, we present new results for the density of anchored clusters in two-dimensional systems at the percolation point.  Our findings deepen and extend the known connections between the percolation point and 2-D electrostatics, and appear to  be generalizable.

\begin{acknowledgments}
We thank J. L. Cardy for useful discussions and K. Dahlberg for assistance with some of the computer simulations.

This work was supported in part by the National Science Foundation Grants Nos. DMR-0203589 and DMS-0244419.
\end{acknowledgments}

\end{document}